\documentclass{Rinton-P9x6}
\usepackage{amsmath}
\usepackage{amsfonts}
\usepackage{amssymb}

\newcommand{\rmd}{\mathrm{d}}
\newcommand{\rmi}{\mathrm{i}}
\newcommand{\rme}{\mathrm{e}}
\newcommand{\Rast}{{\mathbb{R}_*}}
\newcommand{\Tr}{\operatorname{Tr}}
\newcommand{\norm}[1]{\left\Vert#1\right\Vert}

\newcommand{\openone}{{\mathchoice \mathrm{1\mskip-4mu l} \mathrm{1\mskip-4mu l}
\mathrm{1\mskip-4.5mu l} \mathrm{1\mskip-5mu l}}}
\newcommand{\E}{\operatorname{\mathbb{E}}}
\newcommand{\eref}[1]{(\ref{#1})}

\begin{document}

\title{Instrumental processes, entropies, information in quantum continual measurements}

\author{A. Barchielli}

\address{Politecnico di Milano, Dipartimento di Matematica, \\
Piazza Leonardo da Vinci 32, I-20133 Milano, Italy.
\\ E-mail: Alberto.Barchielli@polimi.it}

\author{G. Lupieri}

\address{Universit\`a degli Studi di Milano, Dipartimento di Fisica,\\
Via Celoria 16, I-20133 Milano, Italy.
\\ E-mail: Giancarlo.Lupieri@mi.infn.it}

\maketitle

\begin{center}
 \textbf{Dedicated to Alexander S.~Holevo on his 60th birthday}
 \end{center}

\bigskip

\abstracts{In this paper we will give a short presentation of the quantum L\'evy-Khinchin
formula and of the formulation of quantum continual measurements based on stochastic
differential equations, matters which we had the pleasure to work on in collaboration with
Prof.\ Holevo. Then we will begin the study of various entropies and relative entropies, which
seem to be promising quantities for measuring the information content of the continual
measurement under consideration and for analysing its asymptotic behaviour. }

\section{A quantum L\'evy-Khinchin formula}

The theory of measurements continuous in time in quantum mechanics (quantum continual
measurements) started with the description of counting experiments\cite{Dav76} and of
situations in which an observable is measured imprecisely, but with continuity in
time;\cite{BarLP8283} both formulations are based on the notions of
\emph{instrument}\cite{Dav76} and of positive operator valued measure. Soon after we succeeded
in unifying the two approaches,\cite{BarL85} Holevo\cite{Hol8687} realized that some quantum
analogue of infinite divisibility was involved and thus started a search of a quantum
L\'evy-Khinchin formula;\cite{Hol8789,BarL8990,DenH90,Hol93c} a review is given in
refs.~\citelow{Hol01a,Hol01b}, while a different approach is presented in
refs.~\citelow{BarHL899193}.

Let $\mathcal{H}$ be a complex separable Hilbert space, $\mathcal{T(H)}$ be the trace-class on
$\mathcal{H}$ and $\mathcal{S(H)}$ be the set of statistical operators. We denote by
$\mathcal{L(H}_1;\mathcal{H}_2)$ the space of linear bounded operators from $\mathcal{H}_1$
into $\mathcal{H}_2$ and set $\mathcal{L(H)}=\mathcal{L(H;H)}$. $\langle a, \tau\rangle =
\Tr\{a\tau \}$, $\tau \in \mathcal{T(H)}$, $a \in \mathcal{L(H)}$; $\norm{\tau}_1=
\Tr\left\{\sqrt{\tau^* \tau}\right\}$.

An \emph{instrument} is a map-valued $\sigma$-additive measure $\mathcal{N}$ on some
measurable space $(\mathcal{Y,B})$; the maps are  from $\mathcal{T(H)}$ into itself, linear,
completely positive and normalized in the sense that $\Tr\{\mathcal{N(Y)}
[\tau]\}=\Tr\{\tau\}$.

The formulation of continual measurements given by Holevo\cite{Hol01a} is based on analogies
with the L\'evy processes and it is less general, but more fruitful, than the one initiated by
our group\cite{BarLP8283} and based on the generalized stochastic processes. In order to
simplify the presentation,  we will only consider  the case of one-dimensional processes. Let
$\mathcal{Y}$ be the space of all real functions on the positive time axis starting from zero,
continuous from the right and with left limits, and let $\mathcal{B}_a^b$, $0\leq a \leq b$,
be the $\sigma$-algebra generated by the increments $y(t)-y(s)$, $a\leq s \leq t\leq b$. A
time homogeneous instrumental process with independent increments (\emph{i-process}) is a
family $\{\mathcal{N}_a^b; 0\leq a \leq b\}$, where $\mathcal{N}_a^b$ is an instrument on
$(\mathcal{Y,B}_a^b)$ such that $\mathcal{N}_{a+t}^{b+t}\left(E_t\right)
=\mathcal{N}_{a}^{b}(E)$ for arbitrary $b\geq a$, $t\in \mathbb{R}_+$, $E\in \mathcal{B}_a^b$,
where $E_t= \left\{ y: T_ty\in E\right\}$, $\big(T_ty\big)(s)=y(s+t)$, and such that
 \be\label{composition}
\mathcal{N}_{a}^{b}(E)\circ \mathcal{N}_{b}^{c}(F)=\mathcal{N}_{a}^{c}(E\cap F), \qquad 0\leq
a\leq b \leq c,\quad E\in \mathcal{B}_a^b, \ F\in \mathcal{B}_b^c.
 \ee
Every i-process is determined by its finite-dimensional distributions, which have the
structure
\begin{multline}\label{findimd}
\mathcal{N}_{t_0}^{t_p}\big( y(\cdot): y(t_1)-y(t_0)\in B_1,\ldots , y(t_p)-y(t_{p-1})\in
B_p\big)
\\
=\mathcal{N}_{t_p-t_{p-1}}(B_p)\circ \cdots \circ \mathcal{N}_{t_1-t_{0}}(B_1),
\end{multline}
where $0\leq t_0< t_1< \cdots < t_p$, $B_1,\ldots,B_p\in \mathcal{B}(\mathbb{R})$, and
 \be
\mathcal{N}_{t}(B)= \mathcal{N}_{a}^{a+t}\big(y(\cdot): y(a+t)-y(a)\in B\big)
 \ee
is independent of $a$ by the time homogeneity. The instrument $\mathcal{N}_{t}$ completely
determines the i-process and it is completely characterized by its Fourier transform
(characteristic function) $\int_{\mathbb{R}} \rme^{\rmi k y} \mathcal{N}_t(\rmd y)$;
Eq.~\eref{composition} and the continuity assumption
 \be\label{normcont}
\lim_{t\downarrow 0} \norm{\mathcal{N}_t(U_0) - \openone}=0\,, \quad \text{for every
neighbourhood $U_0$ of $0$,}
 \ee
imply that this characteristic function is of the form $\exp\{t\mathcal{K}(k)\}$,
$\mathcal{K}(k)\in \mathcal{L\big(T(H)\big)}$. The quantum L\'evy-Khinchin formula is the
complete characterization of the generator $\mathcal{K}$.\cite{Hol93c} The structure of
$\mathcal{K}$ can be written in different equivalent ways and here we give an
expression\cite{BarP96} which is particularly convenient for reformulating the theory of the
continual measurements in terms of stochastic differential equations, as illustrated in the
next section.

The quantum L\'evy-Khinchin formula for the generator $\mathcal{K}$ is: $\forall \tau \in
\mathcal{T(H)}$, $\forall k\in \mathbb{R}$, $\forall h,g\in \mathcal{H}$,
\begin{multline}\label{L-Kform}
\mathcal{K}(k)[\tau] = \mathcal{L}[\tau]+ \rmi k c \tau - \frac 1 2 \, r^2 k^2 \tau + \rmi k r
\left( R\tau + \tau R^* \right)
 \\
{}+ \int_{\Rast} \left[ \left( \rme^{\rmi k z}-1 \right) \mathcal{J}[\tau](z) - \rmi k z
\varphi_2(z) \tau \right] \mu(\rmd z)\,,
\end{multline}\nopagebreak
where \ $c\in \mathbb{R}$, \ $r\in \mathbb{R}$, \ $\displaystyle \varphi_2(z) = \frac{b^2}
{b^2+ z^2}$, \ $b>0$,
\begin{eqnarray}\nopagebreak
\mathcal{L}&=&\mathcal{L}_0+ \mathcal{L}_1+\mathcal{L}_2 \,,
 \\
\mathcal{L}_0[\tau] &=& - \rmi [H,\tau] + \frac 1 2 \sum_{j=1}^\infty \left( \left[ L_j \tau,
L_j^* \right] + \left[ L_j, \tau L_j^* \right]\right),
 \\
\mathcal{L}_1[\tau] &=&  \frac 1 2  \left( \left[ R \tau, R^* \right] + \left[ R, \tau R^*
\right]\right),
 \\
\mathcal{L}_2[\tau] &=& -\frac 1 2 J^*J \tau - \frac 1 2 \tau J^*J + \Tr_{{}_{L^2_\nu}}
\left\{ J\tau J^*\right\},
\end{eqnarray}
\begin{equation}\label{defJJ}
\mathcal{J}\left[|h\rangle \langle g| \right](z) = \sum_{n=1}^\infty \frac {\nu\bigl(\rmd z
\times \{n\}\bigr)} {\mu(\rmd z)} \left| ( Jh)(z,n) + h \right\rangle \left \langle ( Jg)
(z,n) +g \right|,
\end{equation}
$R, H, L_j \in \mathcal{L(H)}$, $H=H^*$, $\sum_{j=1}^\infty L_j^*L_j \in  \mathcal{L(H)}$
(strong convergence), $\mathbb{R}_*=\mathbb{R}\backslash \{0\}$, $\nu$ is a $\sigma$-finite
measure on $\mathbb{R}_*\times \mathbb{N}$ and $\mu(\rmd z) = \sum_{n=1}^\infty \nu(\rmd z
\times \{n\})$; we assume that
\begin{equation}
\int_{\Rast} \varphi_1(z)\, \mu(\rmd z) \equiv \sum_{n=1}^\infty \int_{\Rast} \varphi_1(z)
\,\nu\bigl(\rmd z \times \{n\}\bigr) < +\infty\,,
\end{equation}
with \ $\displaystyle \varphi_1(z) = \frac{z^2} {b^2+ z^2}$. Note that
$\varphi_1(z)+\varphi_2(z)=1$ and that $\mu$ is a L\'evy measure on $\Rast$. Finally, $J\in
\mathcal{L}\left(\mathcal{H};L^2_\nu(\mathcal{H})\right)$, where $L^2_\nu = L^2(\Rast\times
\mathbb{N},\nu)$, $L^2_\nu(\mathcal{H})= L^2(\Rast\times \mathbb{N},\nu;\mathcal{H})\simeq
L^2_\nu \otimes \mathcal{H}$. The fact that the operators $H$, $R$, $L_j$, $J$ are bounded is
due to the assumption \eref{normcont}, which is therefore a strong restriction from a physical
point of view.

It is convenient to introduce also the \emph{characteristic functional} of the whole i-process
as the solution of the equation: $\forall a\in \mathcal{L(H)}$, $\forall \tau\in
\mathcal{T(H)}$,
\begin{equation}\label{defG}
\big\langle a,\, \mathcal{G}_t(k)[\tau]\big\rangle = \langle a,\tau\rangle + \int_0^t
\big\langle a,\, \mathcal{K}\big(k(s)\big) \circ \mathcal{G}_s(k)[\tau]\big\rangle \,\rmd s\,,
\end{equation}
where $k(t)$ is a real function, continuous from the left with right limits; let us call it a
\emph{test function}. By taking $k(t)= \kappa\openone_{[0,T)}(t)$, we get $\mathcal{G}_T(k)=
\exp \{T\mathcal{K}(\kappa)\}$ and, similarly, by taking a more general step function for $k$
we get the Fourier transform of the finite-dimensional distributions \eref{findimd}, so that
$\mathcal{G}_t$ completely characterizes the i-process.

The operators $\mathcal{U}(t)= \exp\{t\mathcal{L}\} = \mathcal{G}_t(0)=
\mathcal{N}_t(\mathbb{R})$, $t\geq 0$, form a completely positive quantum dynamical semigroup.
We fix an  \emph{initial state} $\varrho\in \mathcal{S(H)}$ and set $\eta_t=
\mathcal{U}(t)[\varrho]$; $\eta_t$ is called the \emph{a priori state} at time $t$ because it
represents the state of the system at time $t$, when no selection is done on the basis of the
results of the continual measurement. The a priori states satisfy the \emph{master equation}
\begin{equation}
 \frac{\rmd\ }{\rmd t}\,\eta_t =\mathcal{L}[\eta_t]\,, \qquad \eta_0=\varrho\,.
\end{equation}

\section{Stochastic differential equations}
An alternative useful formulation of quantum continual measurements is based on stochastic
differential equations (SDE's); it was initiated for the basic cases by Belavkin\cite{Bel8889}
by using analogies with the classical filtering theory. The general SDE's corresponding to the
L\'evy-Khinchin formula \eref{L-Kform} were studied in refs.~\cite{BarHPZ95}.

\subsection{Output signal and reference probability}

Let $W$ be a one-dimensional standard continuous Wiener process and $N(\rmd z \times \rmd t)$
be a random Poisson measure on $\Rast\times \mathbb{R}_+$ of intensity $\mu(\rmd z)\rmd t$,
independent of $W$. The two processes are realized in a complete standard probability space
$\left( \Omega, \mathcal{F}, Q\right)$ with the filtration of $\sigma$-algebras
$\{\mathcal{F}_t,\; t\geq 0\}$, which is the augmentation of the natural filtration of $W$ and
$N$; we assume also $\mathcal{F}=\bigvee_{t\geq 0} \mathcal{F}_t$. It is useful to introduce
the compensated process
\begin{equation}
\widetilde N(\rmd z \times \rmd t) =  N(\rmd z \times \rmd t) - \mu(\rmd z)\rmd t\,.
\end{equation}
In all the SDE's such as Eqs.~\eref{Lev1}, \eref{Phi}, \eref{eqnnaps}, \eref{PhiG},
\eref{eqaps}, the presence of integrals with respect either to the jump process $N$ or to the
compensated processes $\widetilde N$ or $\breve N$ (see \eref{breveN}) is due to problems of
convergence of the stochastic integrals which arise when infinitely many small jumps are
present (the case $\int_\Rast \mu(\rmd z)=+\infty$).

Now, by using $W$, $N$ and all the ingredients entering the L\'evy-Khinchin formula
\eref{L-Kform}, we are able to construct various random quantities which allow us to reexpress
in a different form the i-process of the previous section. Firstly, let us introduce the real
process
\begin{equation}\label{Lev1}
Y(t)= c t + r W(t) + \int_{\Rast\times (0,t]} \varphi_1(z) z N(\rmd z \times \rmd s) +
\int_{\Rast\times (0,t]} \varphi_2(z) z \widetilde N(\rmd z \times \rmd s) \,,
\end{equation}
which, under the reference probability $Q$, is a generic L\'evy process; $Y$ will represent
the output process of the continual measurement introduced in the previous section. In the
following we shall need the quantity
\begin{equation}
\Phi_t(k)= \exp \biggl\{ \rmi \int_0^t k(s) \rmd Y(s)\biggr\}
\end{equation}
and its stochastic differential
\begin{multline}\label{Phi}
\rmd \Phi_t(k)=\Phi_t(k) \biggl\{ \biggl[  \int_{\Rast} \left( \rme^{\rmi k(t) z} - 1 - \rmi
k(t) \varphi_2(z)z\right)\mu(\rmd z) + \rmi c k(t)
\\ {}- \frac 1 2 \,r^2 k(t)^2 \biggr]\rmd t
+ \rmi r k(t) \rmd W(t) + \int_{\Rast} \left( \rme^{\rmi k(t) z} - 1\right)\widetilde N(\rmd
z\times \rmd t)\biggr\}.
\end{multline}
In Table 1 we summarize the rules of stochastic calculus for $W$ and $N$, which have been used
in computing $\rmd \Phi_t(k)$ and which shall be used to compute all the stochastic
differentials in the rest of the paper.
\begin{table}[h]
\caption{The Ito table and an example of application with only the jump part.}\label{itotable}
\begin{center}
\begin{tabular}{|c||c|c|c|}
\hline
 & $\rmd t$ & $\rmd W(t)$& $N(\rmd z \times \rmd t)$
 \\
 \hline\hline
 $\rmd t$ & 0 & 0 & 0
 \\
 \hline
$\rmd W(t)$& 0 &$\rmd  t$ & 0
\\
\hline $N(\rmd z^\prime \times \rmd t)$ & 0 & 0 & $\delta(z-z^\prime)\,N(\rmd z \times \rmd
t)$
\\ \hline
\end{tabular}
\[ \textstyle
f\left(X+ \int_\Rast C(z) N(\rmd z \times \rmd t)\right) - f(X) = \int_\Rast \left[ f\big(X
+C(z) \big) -f(X) \right]N(\rmd z \times \rmd t)
\]
\end{center}
\end{table}

\subsection{A linear SDE and the instruments}

Let us consider now the linear SDE for $\sigma_t\in \mathcal{T(H)}$, $\sigma_t\geq 0$:
$\forall a \in \mathcal{L(H)}$,
\begin{multline}\label{eqnnaps}
\langle a, \sigma_t \rangle = \langle a, \varrho\rangle + \int_0^t \left\langle a,
\mathcal{L}[\sigma_{s}] \right\rangle \rmd s +\int_0^t \left\langle a,R\sigma_{s}
+\sigma_{s}R^* \right\rangle \rmd W(s)
\\ {} +
\int_{\Rast\times (0,t]} \left\langle  a, \mathcal{J}[\sigma_{s}](z) - \sigma_{s}
\right\rangle \widetilde N(\rmd z\times \rmd s).
\end{multline}
We call the $\sigma_t$ \emph{non normalized a posteriori states} (nnap states); the reason
will be clarified in the following. The coefficient of the jump term should be written as $
\mathcal{J}[\sigma_{s^-}](z) - \sigma_{s^-}$, with the following meaning: when there is a jump
of $N$, i.e.\ when $N(\rmd z\times \rmd s)=1$, the nnap state before the jump $\sigma_{s^-}$
is transformed into the state after the jump $\sigma_{s^+}=\mathcal{J}[\sigma_{s^-}](z)$;
however, we prefer to simplify the notation and not to write the superscripts ``minus''.
Similar considerations apply to all the other SDE's.

By using Table 1 to differentiate $\Phi_t(k)\langle a, \sigma_t \rangle$, we get
\begin{multline}\label{PhiG}
\rmd \big(\Phi_t(k)\langle a, \sigma_t \rangle\big) = \Phi_t(k)\biggr\{\left \langle a,
\mathcal{K}\big(k(t)\big)[\sigma_{t}] \right\rangle \rmd t + \big\langle a, \rmi rk(t)
\sigma_{t} + R \sigma_{t}
\\ {}
+ \sigma_{t} R^* \big\rangle \,\rmd W(t) + \int_{\Rast} \left \langle a, \rme^{\rmi k(t) z}
\mathcal{J}[\sigma_{t}](z) - \sigma_{t}\right\rangle \widetilde N(\rmd z\times \rmd t)\biggr\}
\end{multline}
and by taking the expectation we see that the terms with $\rmd W$ and $\widetilde N$ disappear
and that the resulting equation is the same as Eq.~\eref{defG}, which defines $\mathcal{G}$.
Therefore, we have
\begin{equation}\label{GPhi}
\left\langle a, \mathcal{G}_t(k)[\varrho]\right\rangle = \E_Q\left[ \Phi_t(k)\langle a,
\sigma_t \rangle\right],
\end{equation}
an equation showing that $Y(t)$ and $\sigma_t$ completely determine the characteristic
functional of the continual measurement and, so, the whole i-process. In particular, by taking
$k=0$ we obtain that the expectation value of the nnap states gives the a priori states: $
\E_Q\left[\left\langle a, \sigma_t\right\rangle\right]= \left\langle a, \eta_t\right\rangle$.

\subsection{The physical probability and the a posteriori states}

Let us now study  the norm of the nnap states: $ \norm{\sigma_t}_1 = \langle \openone,
\sigma_t\rangle = \Tr\{\sigma_t\}$. By taking the trace of Eq.~\eref{eqnnaps} we get
\begin{equation}
\rmd \norm{\sigma_t}_1 = \norm{\sigma_{t}}_1 \biggl\{ m(t) \rmd W(t) + \int_{\Rast}
\bigl[\mathcal{I}_t(z)-1 \bigr] \widetilde N(\rmd z \times \rmd t) \biggr\},
\end{equation}
where
\begin{equation}
m(t) = \left\langle R+R^*, \rho_{t}\right\rangle, \qquad \mathcal{I}_t(z) = \left\langle
\openone, \mathcal{J}[\rho_{t}](z)\right\rangle= \left \langle \mathfrak{J}(z),
\rho_t\right\rangle ,
\end{equation}
 \be
 \langle g|\mathfrak{J}(z)h\rangle =
\sum_{n=1}^\infty \frac {\nu\bigl(\rmd z \times \{n\}\bigr)} {\mu(\rmd z)} \left \langle ( Jg)
(z,n) +g | ( Jh)(z,n) + h \right\rangle, \qquad \forall g, h \in \mathcal{H},
 \ee
\begin{equation}\label{defapost}
\rho_t = \begin{cases} \norm{\sigma_t}_1^{\,-1} \sigma_t & \text{if}\ \norm{\sigma_t}_1>0
\\
\varrho & \text{otherwise} \end{cases}
\end{equation}
The operators $\rho_t$ belong to $\mathcal{S(H)}$ and will be called \emph{a posteriori
states}, as explained below. Note the common initial state: $\eta_0=\sigma_0=\rho_0=\varrho$.
It is possible to show that $\norm{\sigma_t(\omega)}_1$ is a martingale and that it can be
used as a local density with respect to $Q$ to define a new probability $P_{\varrho}$ on
$(\Omega,\mathcal{F})$, the \emph{physical probability}, by
\begin{equation}
P_{\varrho}(\rmd \omega )\Big|_{\mathcal{F}_t} = \norm{\sigma_t(\omega)}_1 Q(\rmd
\omega)\Big|_{\mathcal{F}_t}\,, \quad \text{or} \quad \frac{P_{\varrho}(\rmd \omega )} {Q(\rmd
\omega)}\bigg|_{\mathcal{F}_t} =\norm{\sigma_t(\omega)}_1\,.
\end{equation}

By taking $a=\openone$ in \eref{GPhi} and by using the new physical probability we can write
 \be
\left\langle \openone, \mathcal{G}_t(k)[\varrho]\right\rangle = \E_{P_{\varrho}}\left[
\Phi_t(k)\right].
 \ee
This equation shows that the Fourier transform of all the probabilities involved in the
continual measurement is given by the characteristic functional of the process $Y(t)$ under
the probability $P_{\varrho}$. It is this fact which substantiates the interpretation of
$P_{\varrho}$ as the physical probability and of $Y(t)$ as the output process.

It is possible to prove that under the physical probability $P_{\varrho}$
\begin{equation}
\breve W(t) = W(t) - \int_0^t m(s)\,\rmd s
\end{equation}
is a standard Wiener process and $N(\rmd z \times \rmd t)$ is a point process of stochastic
intensity $\mathcal{I}_t(z)\mu(\rmd z)\rmd t$; we set
\begin{equation}\label{breveN}
\breve N(\rmd z \times \rmd t)=N(\rmd z \times \rmd t) - \mathcal{I}_t(z)\mu(\rmd z)\rmd t\,.
\end{equation}
The typical properties of the trajectories of the output signal can be visualized in a
particularly simple manner when $\int_\Rast \varphi_2(z)z\mu(\rmd z)<+\infty$; in this case we
can write
 \be\label{Lev2}
 Y(t)= Y_{\mathrm{cbv}}(t) +r \breve W(t) + \int_{\Rast\times (0,t]}z N(\rmd z \times \rmd
 s)\,
 \ee
where $\int_{\Rast\times (0,t]}z N(\rmd z \times \rmd s)$ is the jump part, with jumps of
amplitude $z$ and intensity $\mathcal{I}_s(z)\mu(\rmd z)\rmd s$, $r \breve W(t)$ is a
continuous part proportional to a Wiener process and
 \be
 Y_{\mathrm{cbv}}(t)= t\bigg( c - \int_{\Rast}  \varphi_2(z ) z\mu(\rmd z)\bigg)+ \int_0^t
 m(s)\rmd s
 \ee
is a continuous part with bounded variation.

By rewriting Eq.~\eref{GPhi} with the new probability, we have
\begin{equation}\label{GPhinew}
\left\langle a, \mathcal{G}_t(k)[\varrho]\right\rangle = \E_{P_{\varrho}}\left[
\Phi_t(k)\langle a, \varrho_t \rangle\right].
\end{equation}
Because $\mathcal{G}$ is the Fourier transform of all the finite-dimensional distributions and
these distributions determine the whole i-process, this last equation is equivalent to: $
\forall a \in \mathcal{L(H)}$, $\forall t\geq 0$, $\forall E\in \mathcal{B}_0^t$,
 \be\label{stocrepr}
 \left\langle a,\, \mathcal{N}_0^t(E)[\varrho]\right\rangle= \int_{\{\omega\in \Omega:
 Y(\cdot;\omega)\in E\}} \langle a, \rho_t(\omega)\rangle P_\varrho(\rmd \omega)\,.
 \ee
This equation shows that $\rho_t$ is the state conditioned on the trajectory of the output
observed up to time $t$ and $\rho_t$ has indeed the meaning of \emph{a posteriori state} at
time $t$: the state we must attribute to the system under a selective measurement up to $t$.
By taking $k=0$ into Eq.~\eref{GPhinew} or $E=\mathcal{Y}$ into Eq.~\eref{stocrepr}, we get
\begin{equation}\label{apostdemix}
\left\langle a, \eta_t\right\rangle = \E_{P_{\varrho}}\left[ \langle a, \rho_t \rangle\right],
\end{equation}
i.e.\ the a posteriori states $\rho_t(\omega) $ with the physical probability $P_\varrho(\rmd
\omega)$ realize a demixture of the a priori state $\eta_t$. Finally, by differentiating the
definition \eref{defapost} of the a posteriori states, we get the SDE
\begin{multline}\label{eqaps}
\langle a, \rho_t \rangle= \langle a, \varrho \rangle + \int_0^t \left\langle a, R\rho_{s} +
\rho_{s} R^* - m(s)\rho_{s} \right\rangle \rmd \breve W(s)
\\{}+
\int_{\Rast\times (0,t]} \left\langle a,  \mathfrak{j}(\rho_{s};z) - \rho_{s}\right\rangle
\breve N(\rmd z \times \rmd s )+ \int_0^t \left\langle a, \mathcal{L} [\rho_{s}] \right\rangle
\rmd s\,,
\end{multline}
\be
 \mathfrak{j}(\tau;z)=\left(\Tr \left\{ \mathcal{J}[\tau](z)\right\} \right)^{-1}\,
 \mathcal{J}[\tau](z)\,, \qquad \tau\in \mathcal{S(H)}\,;
 \ee
Eq.~\eref{eqaps} holds under the physical probability $P_\varrho$.

\section{Entropies and information}

\subsection{Quantum and classical entropies}\label{qce}
In quantum measurement theory both quantum states and classical probabilities are involved
and, so, quantum and classical entropies are relevant.

For $x,y\in \mathcal{T(H)}$, $x\geq 0$, $y\geq 0$, we introduce the functionals, with values
in $[0,+\infty]$,\cite{OhyP93}
\begin{equation}
S_{\mathrm{q}}(x) = - \Tr\{x\ln x\}\,, \qquad S_{\mathrm{q}}(x|y) = \Tr\{x\ln x-x \ln y\}\,;
\end{equation}
if $x,y\in \mathcal{S(H)}$, $S_{\mathrm{q}}(x)$ is the \emph{von Neumann entropy} and
$S_{\mathrm{q}}(x|y)$ is the \emph{quantum relative entropy}. The von Neumann entropy can be
infinite only if the Hilbert space is infinite dimensional and it is zero only on the pure
states, while the quantum relative entropy can be infinite even when the Hilbert space is
finite dimensional and it is zero only if the two states are equal.

A first quantum entropy of interest is the \emph{a priori entropy}
$S_{\mathrm{q}}\big(\eta_t\big)$, which at time zero reduces to the entropy of the initial
state $S_{\mathrm{q}}\big(\eta_0\big)=S_{\mathrm{q}}(\varrho)$.

On the other hand, a classical entropy is the relative entropy (or Kullback-Leibler
informational divergence) of the physical probability $P_\varrho$ with respect to the
reference probability measure $Q$:
\begin{equation}
I_t(P_{\varrho}|Q) = \E_{P_{\varrho}}\bigg[\ln\,\frac{P_{\varrho}(\rmd \omega )} {Q(\rmd
\omega)}\bigg|_{\mathcal{F}_t} \bigg] = \E_Q\big[ \norm{\sigma_t}_1 \ln
\norm{\sigma_t}_1\big]\,,
\end{equation}
Let us note that $I_t(P_{\varrho}|Q)\geq 0$, $ I_0(P_{\varrho}|Q)=0$ and that $
I_t(P_{\varrho}|Q)$ is non decreasing, as one sees by computing its time
derivative:
\begin{equation}
\frac {\rmd\ }{\rmd t}\,I_t(P_{\varrho}|Q) = \E_{P_{\varrho}}\biggl[\frac 1 2 \, m(t)^2+
\int_{\Rast} \bigl( 1-\mathcal{I}_t(z) +\mathcal{I}_t(z)\ln \mathcal{I}_t(z)\bigr)\mu(\rmd
z)\biggr] \geq 0\,.
\end{equation}

If we consider two different initial states $\varrho^\alpha$ and $\varrho$, with
$\mathrm{supp}\, \rho^\alpha \subseteq \mathrm{supp}\, \rho$, we can introduce the quantum
relative entropy $S_{\mathrm{q}}(\eta_t^\alpha|\eta_t)$ and the \emph{classical
$P_{\varrho^\alpha}|P_{\varrho}$-relative entropy} $I_t(P_{\varrho^\alpha}|P_{\varrho})$,
\begin{equation}
I_t(P_{\varrho^\alpha}|P_{\varrho}) =
\E_{P_{\varrho^\alpha}}\bigg[\ln\,\frac{P_{\varrho^\alpha}(\rmd \omega )} {P_{\varrho}(\rmd
\omega)}\bigg|_{\mathcal{F}_t} \bigg] = \E_Q\bigg[ \norm{\sigma_t^\alpha}_1 \ln
\frac{\norm{\sigma_t^\alpha}_1} {\norm{\sigma_t}_1}\bigg]\,.
\end{equation}
Here and in the following $P_{\varrho^\alpha}$, $ \sigma_t^\alpha$, $\rho_t^\alpha$,
$\eta_t^\alpha$, $m^\alpha(t)$, $\mathcal{I}_t^\alpha(z)$ are defined by starting from
$\varrho^\alpha$ as $P_{\varrho}$, $ \sigma_t$, $\rho_t$, $\eta_t$, $m(t)$, $\mathcal{I}_t(z)$
are defined by starting from $\varrho$.

Let us stress the different behaviour in time of the two relative entropies; this discussion
will be relevant later on. The quantum one starts from
$S_{\mathrm{q}}(\varrho^\alpha|\varrho)$ at time zero and it is non increasing
\begin{equation}\label{relentrdecreasing}
S_{\mathrm{q}}\left(\eta_t^\alpha\big|\eta_t\right) =
S_{\mathrm{q}}\left(\mathcal{U}(t-s)\left[\eta_s^\alpha\right]\big|
\mathcal{U}(t-s)\left[\eta_s\right]\right) \leq
S_{\mathrm{q}}\left(\eta_s^\alpha\big|\eta_s\right)\,, \qquad t>s\,;
\end{equation}
this statement follows from the Uhlmann monotonicity theorem (ref.~\cite{OhyP93} Theor.\ 5.3).
The classical relative entropy starts from zero at time zero and it is non decreasing, as one
sees by computing its time derivative
\begin{multline}\label{clinfdot}
\frac {\rmd\ }{\rmd t}\,I_t(P_{\varrho^\alpha}|P_{\varrho})=
\E_{P_{\varrho^\alpha}}\biggl[\frac 1 2 \,\big( m^\alpha(t)-m(t)\big)^2
\\ {}
+ \int_{\Rast} \left( 1-\frac {\mathcal{I}_t^\alpha(z)}{\mathcal{I}_t(z)}
+\frac{\mathcal{I}_t^\alpha(z)}{\mathcal{I}_t(z)}\, \ln
\frac{\mathcal{I}_t^\alpha(z)}{\mathcal{I}_t(z)}\right)\mathcal{I}_t(z)\mu(\rmd z)\biggr] \geq
0\,.
\end{multline}
However, both relative entropies have the same bounds:
 \be\label{2bounds}
0\leq S_{\mathrm{q}}(\eta_t^\alpha|\eta_t) \leq S_{\mathrm{q}}(\varrho^\alpha|\varrho)\,,
\qquad 0\leq I_t(P_\varrho^\alpha|P_{\varrho}) \leq S_{\mathrm{q}}(\varrho^\alpha|\varrho)\,.
 \ee
The first statement is clear [see Eq.~\eref{relentrdecreasing}]. The second one too is a
consequence of the Uhlmann monotonicity theorem, as can be seen by considering the
``observation channel'' $\Lambda: \mathcal{L(H)} \to L^\infty(\Omega,\mathcal{F}_t, Q)$ with
predual $\Lambda^*: \varrho \to P_\varrho \in L^1(\Omega,\mathcal{F}_t, Q)$ (in
ref.~\cite{OhyP93} see p.\ 138, Theor.\ 5.3 and the discussions at pgs.\ 9 and 151).

\subsection{Entropies and purity of the states}\label{eps}

When one is studying the properties of an instrument, a relevant question is whether the a
posteriori states are pure or not and, if not pure, how to measure their ``degree of mixing''.
Ozawa\cite{Oza86} called \emph{quasi-complete} an instrument which sends every initial pure
state into pure a posteriori states. A first measure of purity of the a posteriori states is
the \emph{a posteriori entropy} $\E_{P_{\varrho}}\big[ S_{\mathrm{q}}\big(\rho_t\big)\big]$,
which takes the initial value $\E_{P_{\varrho}}\big[
S_{\mathrm{q}}\big(\rho_0\big)\big]=S_{\mathrm{q}}(\varrho)$. A related quantity, simpler to
study, is the \emph{a posteriori purity} (or linear entropy)
 \be
 p(t)=\E_{P_\varrho}
\left[\Tr \left\{\rho_t \left(\openone - \rho_t\right)\right\} \right],\qquad  p(0)=\Tr
\left\{\varrho \left(\openone - \varrho\right)\right\}.
 \ee
The a posteriori entropy and purity vanish if and only if the a posteriori states are almost
surely pure and one has $p(t)\leq \E_{P_{\varrho}}\big[ S_{\mathrm{q}}\big(\rho_t\big)\big]$.

By the rules of stochastic calculus (Table \ref{itotable}) we get the time derivative of the
purity
\begin{gather}
\frac{\rmd\ }{\rmd t} \, p(t)= \dot p_1(t) - \dot p_2(t) - \dot p_3(t)\,,
\\
\dot p_1(t) = 2\sum_{j=1}^\infty \E_{P_\varrho} \left[ \Tr \left\{ \rho_t L_j^* L_j \rho_t -
\rho_t^{1/2} L^*_j \rho_t L_j \rho_t^{1/2} \right\}\right],
\\
\dot p_2(t) = \E_{P_\varrho} \left[ \Tr \left\{ \rho_t^{1/2}\left( R+R^* -m(t)\right) \rho_t
\left( R+R^* -m(t)\right) \rho_t^{1/2} \right\}\right]\geq 0,
\\
\begin{split}
\dot p_3(t) &= \int_\Rast \E_{P_\varrho} \left[ \Tr \left\{ \mathcal{I}_t(z) \,
\mathfrak{j}(\rho_t;z)^2- 2 \mathcal{J}[\rho_t^{\,2}](z) +\mathcal{I}_t(z)\rho_t^{\,2}
\right\}\right]\mu(\rmd z)
\\ {}
&= \int_\Rast \E_{P_\varrho} \Big[\mathcal{I}_t(z)^{-1} \Tr \Big\{  \left( \rho_t^{\,1/2}
\mathfrak{J}(z)\rho_t^{\,1/2}-  \mathcal{I}_t(z) \rho_t \right)^2
\\ &{}\qquad\ \qquad {}+
\mathcal{J}[\rho_t](z)^2 - \left( \rho_t^{\,1/2}
\mathfrak{J}(z)\rho_t^{\,1/2}\right)^2\Big\}\Big]\mu(\rmd z) \,.
\end{split}
\end{gather}
Then, one can check the following points.
\begin{itemize}
\item[(a)] If $\rho_t$ is almost surely a pure state, then one has $\dot p_1(t)\geq0$, $\dot
p_2(t)=0$, $\dot p_3(t)=-\int_\Rast \E_{P_\varrho} \left[ \Tr \left\{
\mathfrak{j}(\rho_t;z)-\mathfrak{j}(\rho_t;z)^2 \right\}\mathcal{I}_t(z)\right]\mu(\rmd z)\leq
0$.
\item[(b)] The a posteriori states are almost surely pure for all pure initial
states (i.e.\ the measurement is quasi complete) if and only if the following conditions hold:
\begin{description}
\item[C1.] $\mathcal{L}_0[\cdot] = -\rmi [H,\cdot]$;
\item[C2.] $\mathfrak{j}(\tau;z)$ is a
pure state ($\mu$-almost everywhere) for all pure states $\tau$ or, equivalently, in
\eref{defJJ} $\big(Jh\big)(z,n)=\big(Jh\big)(z)$, $\forall h\in\mathcal{H}$.
\end{description}
\item[(c)] Under the same conditions one has
$\dot p_1(t)=0$,  $\dot p_3(t) \geq 0$ for any initial state; as $\dot p_2(t) \geq 0$ always,
the purity decreases monotonically.
\end{itemize}

The properties of the purity have also been used\cite{BarP03} to find sufficient conditions
(among which there is the quasi-completeness property) so that the long time limit of the a
posteriori purity will vanish for every initial state; note that in a finite dimensional
Hilbert space this is equivalent to the vanishing of the limit of the a posteriori entropy.

Differentiating the a posteriori entropy demands long computations involving an integral
representation of the logarithm (ref.~\cite{OhyP93} p.~51) and the rules of stochastic
calculus. We get
\begin{equation}\label{apostedot}
\frac{\rmd \ }{\rmd t}\,\E_{P_{\varrho}}\big[S_{\mathrm{q}}\big(\rho_t\big)\big]=
\E_{P_{\varrho}}[D_1(\rho_t) - D_2(\rho_t) - D_3(\rho_t)]\,,
\end{equation}
where, $\forall \tau\in \mathcal{S(H)}$,
 \begin{align}
 D_1(\tau) &= \sum_j \Tr \left\{ \left( L_j^* L_j \tau - L_j \tau L_j^* \right) \ln \tau
 \right\},
\\ \label{D2}
\begin{split}
D_2(\tau )&= \int_0^{+\infty} \rmd u \,\Tr \bigg\{ \frac {u\tau}{(u+\tau)^2} \left( R+R^* -
\Tr \left\{ \left( R+R^*\right) \tau \right\} \right) \frac{\tau}{u+\tau}
\\ &{}\ \ {} \times
\left( R+R^* - \Tr \left\{ \left( R+R^*\right) \tau \right\} \right) + \frac
{\tau}{(u+\tau)^2}\,[\tau,R]\,\frac{\tau}{u+\tau}\,R^*
\\ & \qquad \qquad {}
-\biggl[ \frac{\tau}{u+\tau}\,, R\biggr] \frac{\tau}{u+\tau}\, R^*\bigg\},
\end{split}
 \\
D_3(\tau) &=\int_{\Rast} \mu(\rmd z) \left( \Tr \left\{ - \mathcal{J}[\tau\ln \tau](z)\right\}
- \Tr\left\{ \mathcal{J}[\tau](z)\right\} S_{\mathrm{q}}\big(
\mathfrak{j}(\tau;z)\big)\right).
 \end{align}
From the time derivative of the a posteriori entropy we have the following results.
\begin{itemize}
\item[(i)] When $\tau$ is a pure state, $D_1(\tau)=0 $ if
$\sum_j \Tr \left\{ \tau L_j^* \left( \openone -\tau \right) L_j \right\} = 0$ and
$D_1(\tau)=+\infty$ otherwise.
\item[(ii)] $D_2(\tau)\geq 0$ for any state $\tau$. When $\tau $ is a pure state
$D_2(\tau)=0 $.
\item[(iii)] Under condition \textbf{C2} one has
$D_3(\tau) \geq 0$ for any state $\tau$.
\item[(iv)] When $\tau$ is a pure state, $D_3(\tau) \leq 0$ in general and
$D_3(\tau)=0$ if condition \textbf{C2} holds.
\end{itemize}
Statements (i) and (iv) are easy to verify, while the proof of (iii) requires arguments
introduced in Section \ref{meai} and will be given there. In order to study $D_2(\tau)$ we
need the spectral decomposition of $\tau$: $\tau = \sum_k \lambda_k P_k$, with $k\neq r
\Rightarrow \lambda_k \neq \lambda_r$; by inserting this decomposition into Eq.~\eref{D2} we
get
\begin{multline}
D_2(\tau) = \frac 1 2 \sum_k \lambda_k \Tr \left\{ \left[ P_k \left( R+R^* - \Tr \left\{
\left( R+R^*\right) \tau \right\} \right)P_k\right]^2\right\}
\\ {}
+ \frac 1 2 \sum_{k\neq r} \Tr \left\{ P_k (R+R^*) P_r (R+R^*) P_k\right\}\frac{\lambda_k
\lambda_r}{\lambda_k-\lambda_r} \, \ln \frac {\lambda_k}{\lambda_r}\,,
\end{multline}
which implies statement (ii).

\subsection{Mutual entropies and amount of information}\label{meai}

A basic concept in classical information theory is the mutual entropy (information). For two
nonindependent random variables it is the relative entropy of their joint probability
distribution with respect to the product of the marginal distributions and it is a measure of
how much information  the two random variables have in common. The idea of mutual entropy can
be introduced also in a quantum context, when tensor product structures are involved. Ohya
used the quantum mutual entropy in order to describe the amount of information correctly
transmitted through a quantum channel $\Lambda^*$ from an input state $\varrho$ to the output
state $\Lambda^* \varrho$. The starting point is the definition of a ``compound state'' which
describes the correlation of $\varrho$ and $\Lambda^* \varrho$; it depends on how one
decomposes the input state $\varrho$ in elementary events (orthogonal pure states). The
\emph{mutual entropy of the state $\varrho$ and the channel $\Lambda^*$} is then defined as
the supremum over all such decompositions of the relative entropy of the compound state with
respect to the product state $\varrho\otimes \Lambda^* \varrho$ (ref.~\cite{OhyP93}
pp.~33--34, 139).

We want to generalize these ideas to our context, where we have not only a quantum channel
$\mathcal{U}(t)$, but also a classical output with probability law $P_\varrho$; let us note
that $\sigma_t$ contains the a posteriori states and the probability law and that it can be
identified with a state on $\mathcal{L(H)}\otimes L^\infty(\Omega, \mathcal{F}_t, Q)$.
Firstly, we define a compound state $\Sigma_t$ describing the correlation between the initial
state $\varrho$ and the nnap state $\sigma_t$. Let $\varrho= \sum_\alpha w_\alpha
\varrho^\alpha$ be a decomposition of the initial state into orthogonal pure states (an
extremal Shatten decomposition); if $\varrho$ has degenerate eigenvalues, this decomposition
is not unique. With the notations of Section \ref{qce} we have
\begin{equation}
\begin{split}
\sigma_t= \sum_\alpha w_\alpha\sigma_t^\alpha\,, \ \qquad \rho_t = \sum_\alpha w_\alpha
\,\frac {\norm{\sigma_t^\alpha}_1} {\norm{\sigma_t}_1}\, \rho_t^\alpha\,, \ \qquad \eta_t=
\sum_\alpha w_\alpha\eta_t^\alpha\,,
\\
P_{\varrho}=\sum_\alpha w_\alpha P_{\varrho^\alpha}\,, \qquad \sum_\alpha w_\alpha
\rho_t^\alpha(\omega) P_{\varrho^\alpha}(\rmd \omega)\Big|_{\mathcal{F}_t} = \rho_t(\omega)
P_{\varrho}(\rmd \omega)\Big|_{\mathcal{F}_t}\,.
\end{split}
\end{equation}
The compound state $\Sigma_t$ will be a state on the von Neumann algebra
$\mathfrak{A}=\mathcal{L(H)}\otimes \mathcal{L(H)} \otimes L^\infty(\Omega, \mathcal{F}_t, Q)
\equiv \mathcal{M}_1 \otimes \mathcal{M}_2 \otimes \mathcal{M}_3 $; a normal state $\Sigma$ on
$\mathfrak{A}$ is represented by a non negative random trace-class operator $\widehat
{\Sigma}$ on $\mathcal{H}\otimes \mathcal{H}$ such that $\int_\Omega \Tr_{\mathcal{H}\otimes
\mathcal{H}}\left\{\widehat{\Sigma}(\omega)\right\} Q(\rmd \omega)=1$: $\Sigma(A)= \int_\Omega
\Tr_{\mathcal{H}\otimes \mathcal{H}}\left\{ \widehat{\Sigma}(\omega) A(\omega)\right\} Q(\rmd
\omega)$, $A\in \mathfrak{A}$.  The relative entropy of the state $\Sigma$ with respect to
another state $\Pi$ with representative $\widehat{\Pi}$ is given by
 \be
 S(\Sigma|\Pi)= \int_\Omega \Tr_{\mathcal{H}\otimes
\mathcal{H}}\left \{\widehat{\Sigma}(\omega) \left(\ln \widehat{\Sigma}(\omega) - \ln
\widehat\Pi(\omega)\right)\right\} Q(\rmd \omega)\,;
 \ee
this formula is consistent with the general Araki-Uhlmann definition of relative entropy in a
von Neumann algebra (ref.~\cite{OhyP93} Chapt.~5).

We introduce the compound state $\Sigma_t$ on $\mathfrak{A}$ by giving its representative
$\sum_\alpha w_\alpha \varrho^\alpha \otimes \sigma_t^\alpha$ and we consider the different
possible product states which can be constructed with its marginal: $\Pi_t=
\Sigma_t\big|_{\mathcal{M}_1}\otimes \Sigma_t\big|_{\mathcal{M}_2}\otimes
\Sigma_t\big|_{\mathcal{M}_1}$ with representative $\norm{\sigma_t}_1\, \varrho \otimes
\eta_t$, $\Pi_t^1=\Sigma_t\big|_{\mathcal{M}_1}\otimes \Sigma_t\big|_{\mathcal{M}_2\otimes
\mathcal{M}_3}$ with representative $\varrho \otimes \sigma_t$,
$\Pi_t^2=\Sigma_t\big|_{\mathcal{M}_2}\otimes \Sigma_t\big|_{\mathcal{M}_1\otimes
\mathcal{M}_3}$ with representative $\sum_\alpha w_\alpha \norm{\sigma_t^\alpha}_1\,
\varrho^\alpha \otimes \eta_t$, $\Pi_t^3= \Sigma_t\big|_{\mathcal{M}_1\otimes
\mathcal{M}_2}\otimes \Sigma_t\big|_{\mathcal{M}_3}$ with representative $
\norm{\sigma_t}_1\sum_\alpha w_\alpha \varrho^\alpha \otimes \eta_t^\alpha$. The different
mutual entropies, i.e.\ the relative entropies of $\Sigma_t$ with respect to the different
product states, are the object of interest. We can call $ S(\Sigma_t|\Pi_t)$ the \emph{mutual
input/output entropy}; this is a new informational quantity, which could be extended also to
generic measurements represented by instruments. First of all, from Corollary 5.20 of
ref.~\cite{OhyP93}, we obtain the chain rule
 \be\label{sp}
 S(\Sigma_t|\Pi_t)= S(\Sigma_t|\Pi_t^i)+S(\Pi^i_t|\Pi_t)\,, \qquad i=1,2,3\,.
 \ee
 Then, with some computations, we obtain the following relations:
\begin{align}\label{p1p}
S(\Pi^1_t|\Pi_t)&= \E_{P_\varrho}\left[ S_{\mathrm{q}}(\rho_t|\eta_t)\right] =
S_{\mathrm{q}}(\eta_t) - \E_{P_\varrho}\left[ S_{\mathrm{q}}(\rho_t)\right],
\\ \label{p2p}
S(\Pi^2_t|\Pi_t)&= \sum_\alpha w_\alpha\, I_t(P_{\varrho^\alpha}|P_\varrho) = \sum_\alpha
w_\alpha\, I_t(P_{\varrho^\alpha}|Q)-  I_t(P_{\varrho}|Q)\,,
\\ \label{p3p}
S(\Pi^3_t|\Pi_t)&= \sum_\alpha w_\alpha\,  S_{\mathrm{q}}(\eta^\alpha_t|\eta_t) =
S_{\mathrm{q}}(\eta_t)- \sum_\alpha w_\alpha\, S_{\mathrm{q}}(\eta_t^\alpha) \,;
\end{align}
\begin{align} \label{sp1}
\begin{split}
S(\Sigma_t|\Pi_t^1)&= S(\Pi^2_t|\Pi_t)+ \sum_\alpha w_\alpha\,\E_{P_{\varrho^\alpha}}
\left[ S_{\mathrm{q}}(\rho_t^\alpha|\rho_t)\right]
\\ {}&=
S(\Pi^2_t|\Pi_t)+\E_{P_\varrho}\left[ S_{\mathrm{q}}(\rho_t)\right]- \sum_\alpha w_\alpha\,
\E_{P_{\varrho^\alpha}}\left[ S_{\mathrm{q}}(\rho_t^\alpha)\right],
\end{split}
\\ \label{sp2}
S(\Sigma_t|\Pi_t^2)&=\sum_\alpha w_\alpha\,\E_{P_{\varrho^\alpha}}\left[ S_{\mathrm{q}}(\rho_t^\alpha|\eta_t)\right]
= S_{\mathrm{q}}(\eta_t)- \sum_\alpha w_\alpha\,\E_{P_{\varrho^\alpha}}\left[ S_{\mathrm{q}}(\rho_t^\alpha)\right],
\\ \label{sp3}
\begin{split}
S(\Sigma_t|\Pi_t^3)&=S(\Pi^2_t|\Pi_t)+ \sum_\alpha w_\alpha\,
\E_{P_{\varrho^\alpha}}\left[ S_{\mathrm{q}}(\rho_t^\alpha|\eta^\alpha_t)\right]
\\ {}&=
S(\Pi^2_t|\Pi_t)+\sum_\alpha w_\alpha \, S_{\mathrm{q}}(\eta_t^\alpha)- \sum_\alpha w_\alpha\,
\E_{P_{\varrho^\alpha}}\left[ S_{\mathrm{q}}(\rho_t^\alpha)\right];
\end{split}
\end{align}
\be \label{spfinal}
 S(\Sigma_t|\Pi_t)=S(\Pi^2_t|\Pi_t)+S_{\mathrm{q}}(\eta_t)-\sum_\alpha w_\alpha\,
 \E_{P_{\varrho^\alpha}}\left[ S_{\mathrm{q}}(\rho_t^\alpha)\right].
 \ee
The initial values are
 \be
 \begin{split}
&S(\Sigma_0|\Pi_0)= S(\Sigma_0|\Pi_0^1)=
S(\Sigma_0|\Pi_0^2)=S(\Pi_0^3|\Pi_0)=S_{\mathrm{q}}(\varrho)\,,
\\
&S(\Sigma_0|\Pi_0^3)= S(\Pi_0^1|\Pi_0)=S(\Pi_0^2|\Pi_0)=0\,.
\end{split}
 \ee

The quantity $S(\Pi^1_t|\Pi_t)=\E_{P_\varrho}\left[ S_{\mathrm{q}}(\rho_t|\eta_t)\right]$ is
the \emph{a posteriori relative entropy};\cite{Bar01} because Eq.~\eref{apostdemix} can be
interpreted by saying that $\{P_\varrho(\rmd \omega),\, \rho_t(\omega)\}$ is a demixture of
the a priori state $\eta_t$, such a relative entropy is a measure of how much such a demixture
is fine. Let us observe that, for $s\leq t$,
 \be
 \E_{P_\varrho}\left[ S_{\mathrm{q}}(\rho_t|\eta_t)\right]=
 \E_{P_\varrho}\left[ S_{\mathrm{q}}(\rho_t|\mathcal{U}(t-s)[\rho_s])\right]+
 \E_{P_\varrho}\left[ S_{\mathrm{q}}(\mathcal{U}(t-s)[\rho_s]|\eta_t)\right].
 \ee
It follows that the variation in time of the a posteriori entropy is the sum of two competing
contributions of opposite sign:
\begin{multline}
\Delta \E_{P_\varrho}\left[ S_{\mathrm{q}}(\rho_t|\eta_t)\right]= \E_{P_\varrho}\left[
S_{\mathrm{q}}(\rho_{t+\Delta t}|\mathcal{U}(\Delta t)[\rho_t])\right]
\\ {}+ \left\{
\E_{P_\varrho}\left[ S_{\mathrm{q}}(\mathcal{U}(\Delta t)[\rho_t]|\mathcal{U}(\Delta t)[
\eta_t])\right]- \E_{P_\varrho}\left[ S_{\mathrm{q}}(\rho_t|\eta_t)\right] \right\}.
\end{multline}
The first term is clearly positive and represents an information gain due to the process of
demixture induced by the measurement. The second term is negative, once again as a consequence
of the Uhlmann monotonicity theorem, and represents an information loss due to the partial
lack of memory of the initial state induced by the dissipative part of the dynamics.

The quantity $S(\Pi^2_t|\Pi_t)=\sum_\alpha w_\alpha\, I_t(P_{\varrho^\alpha}|P_\varrho) $ has
been introduced by Ozawa\cite{Oza86} for a generic instrument under the name of
\emph{classical amount of information}. By the discussion in Section \ref{qce}, eqs.\
\eref{clinfdot} and \eref{2bounds}, one obtains that this quantity is non decreasing and
bounded:
\begin{equation}
0\leq S(\Pi^2_t|\Pi_t) =\sum_\alpha w_\alpha\, I_t(P_{\varrho^\alpha}|P_\varrho)\leq
\sum_\alpha w_\alpha S_{\mathrm{q}}(\varrho^\alpha|\varrho)=S_{\mathrm{q}}(\varrho)\,.
\end{equation}

The supremum over all extremal Shatten decompositions of $S(\Pi^3_t|\Pi_t)= \sum_\alpha
w_\alpha\, S_{\mathrm{q}}(\eta^\alpha_t|\eta_t)$ is Ohya's  ``mutual entropy of the input
state $\varrho$ and the channel $\mathcal{U}(t)$''; by \eref{relentrdecreasing}
$S(\Pi^3_t|\Pi_t)$ is non increasing and by Theor.~1.19 of ref.~\cite{OhyP93} it is bounded by
 \be
 0\leq  S(\Pi^3_t|\Pi_t)= \sum_\alpha w_\alpha\, S_{\mathrm{q}}(\eta^\alpha_t|\eta_t)
  \leq \min\left\{S_{\mathrm{q}}(\varrho),\, S_{\mathrm{q}}(\eta_t)\right\}\,.
 \ee

For general instruments Ozawa\cite{Oza86} introduced an entropy defect, which he called the
\emph{amount of information}; it measures how much the a posteriori states are purer  than the
initial state (or less pure, when this quantity is negative). In the case of continual
measurements it is defined by\cite{Bar01}
\begin{equation}
I_t(\varrho)= S_{\mathrm{q}}(\varrho) - \E_{P_{\varrho}}\big[
S_{\mathrm{q}}\big(\rho_t\big)\big].
\end{equation}
If an equilibrium state exists,  $\eta_{\mathrm{eq}}\in \mathcal{S(H)}$ and
$\mathcal{L}[\eta_{\mathrm{eq}}]=0$, by \eref{p1p} we have
$S_{\mathrm{q}}(\eta_{\mathrm{eq}})\geq I_t(\eta_{\mathrm{eq}})= \E_{P_{\eta_{\mathrm{eq}}}}[
S_{\mathrm{q}} (\rho_t | \eta_{\mathrm{eq}})] \geq 0$. For a \textbf{quasi-complete} continual
measurement one has
\begin{equation}
S_{\mathrm{q}}(\varrho)\geq I_t(\varrho)\geq S(\Pi^2_t|\Pi_t)\geq 0\,, \qquad I_t(\varrho)\geq
I_s(\varrho)\,, \quad t\geq s\,.
\end{equation}
The first statement was proved by Ozawa\cite{Oza86} for a generic quasi-complete instrument,
while the second one follows from the first one by using conditional expectations.\cite{Bar01}
We have $I_t(\varrho)- I_s(\varrho)= \mathbb{E}_{P_\varrho}\big[ S_{\mathrm{q}}(\rho_s)
-\mathbb{E}_{P_\varrho}[S_{\mathrm{q}}(\rho_t)|\mathcal{F}_s]\big]$; but
$S_{\mathrm{q}}(\rho_s) -\mathbb{E}_{P_\rho}[S_{\mathrm{q}}(\rho_t)|\mathcal{F}_s]$ is the
amount of information at time $t$ when the initial time is $s$ and the initial state is
$\rho_s$ and, so, it is non-negative for a quasi-complete measurement. From the monotonicity
of $I_t(\varrho)$ one obtains that the time derivative of $\E_{P_{\varrho}}\big[
S_{\mathrm{q}}\big(\rho_t\big)\big]$ is negative and this holds in particular at time zero for
any choice of the initial state and also for $R=0$. This proves the statement (iii) of Section
\ref{eps}.

\section*{Acknowledgments}
Work supported in part by the \ \emph{European Community's Human Potential Programme} \ under
contract HPRN-CT-2002-00279, QP-Applications, \ and by \emph{\ Istituto Nazionale di Fisica
Nucleare, Sezione di Milano}.

\end{document}